# A Review on Failure Node Recovery Algorithms in Wireless Sensor Actor Networks

G.Sumalatha[#1], N.Zareena[*2], Ch.Gopi Raju[#3]

[#1&2]*Asst.Prof, Department of CSE, Vignan's LARA Institute Of Technology & Science, Vadlamudi Guntur Dist., A.P., India*

[#3]*Mtech(perusing),Department of CSE, Vignan's LARA Institute Of Technology & Science, Vadlamudi Guntur Dist., A.P., India.*

*Abstract—* **In wireless sensor-actor networks, sensors probe their surroundings and forward their data to actor nodes. Actors collect sensor data and perform certain tasks in response to various events. Since actors operate on harsh environment, they may easily get damaged or failed. Failed actor nodes may partition the network into disjoint subsets. In order to re-establish connectivity nodes may be relocated to new positions. This paper focus on review of three (LeDir, RIM, DARA) node recovery algorithms, and their performance has been analysed in terms network overhead and path length validation metrics.**

## I. INTRODUCTION

Wireless Sensor Actor Networks play a vital role in the applications of interest such as remote and harsh areas in which human intervention is risky or impractical. Examples include space exploration, battle field surveillance, search-and-research, and coastal and border protection. A typical WSAN consists of a larger set of miniaturized sensor nodes probe their surroundings and report them to actor (actuator) nodes which collects reports and responding to particular events of interest. For example, sensors may detect a fire and trigger a response from an actor that has an extinguisher. Robots and unmanned vehicles are example actors in practice [1]. Actors need to work collaboratively to meet the application mission; a strongly connected inter-actor network topology would be required at all times. However, a failure of an actor may cause the network to partition into disjoint blocks and would thus violate such a connectivity requirement. As actors deployed in harsh environment it is difficult to replace actor nodes, so we need to reposition actor nodes [2]. In addition, we need to maintain distributed recovery since the nodes cannot be reestablish network connectivity, and their performance analyzed with respect to network overhead and path length validation metrics.

## II. RELATED WORK

A number of schemes have recently been proposed for restoring network connectivity in partitioned WSANs [2]. Some schemes replaces failed nodes with additional rely nodes, where as others carefully reposition the nodes in order to maintain network connectivity. Our focus is on repositioning of nodes to restore the connectivity.

## III. NODE RECOVERY ALGORITHMS

We analyzed three types of recovery algorithm

### A. Recovery through Inward Motion

RIM [4] follows distributed recovery with no coordination among nodes. Nodes can decide independently when to start restoration process where to move. For that it maintains 1-hop neighboring information to start the recovery process.

At network setup, each node broadcasts a *HELLO* message to introduce itself to its neighbors, then builds a list of directly reachable neighbor nodes called 1-hop neighbors. The 1-hop neighbors table is maintained during network operation to reflect changes in the topology. Each table entry contains two parameters: {Node_ID, Relative position}. Nodes inform their neighbors before changing their position

Nodes will periodically send heartbeat messages to their neighbors to ensure that they are functional. RIM starts the recovery process when heart beat messages are missing from any node referred as $N_f$. 1-Hop neighbours of $N_f$ start recovery process moving towards the position of $N_f$ up to a distance of $r/2$ from $N_f$ until they reach others in the network.

### B. Distributed Actor Recovery Algorithm

DARA, a Distributed Actor Recovery Algorithm [3], is localized scheme that avoids the involvement every single actor in the network. Which works efficiently restore the connectivity of an inter actor network to its pre-node-failure level. DARA implemented to address two types of connectivity to maintain 1-hop neighbors and 2-hop neighbors such as 1-connectivity and 2-connectivity.

The main idea of DARA-1C is to replace the dead actor by a suitable neighbor. The selection of the best candidate (BC) neighbor is based on the node degree and the physical proximity to the dead actor. The relocation procedure is recursively applied to handle any actors that get disconnected due to the movement of one of their neighbors (e.g., the BC that replaced the faulty actor). Similarly, DARA-2C identifies the nodes that are affected, i.e., lost their 2-connectivity property, due to the failed actor. Some of these nodes are then relocated in order to restore 2-connectivity. Although both DARA-1C and DARA-2C pursue node relocation to restore the desired level of connectivity, they fundamentally differ in the scope of the failure analysis and the recovery.

When a failed node disconnect network into partitions, the neighbor of the failed node will take the lead and move toward the location of the failed node. The other nodes in the partition follow through in the same direction headed by the leader node and maintain their current links. Cascaded relocation takes less number of node movements compared to entire block movement In addition, block movement requires all the actors in the sub network to be aware of where and how far to move, which introduces extra messaging.

1*) DARA 1-C.*

DARA 1-C initiates the recovery process with the neighbors of the failed node. The following are the detailed steps of DARA 1-C.

1.1) Heartbeats and Neighbor List Maintenance:

DARA 1-C requires that each actor in the network should keep its neighbors information. Neighbors list should be updated each time when they change their position. Neighbors information maintained in the form table consist of three parameters: {node degree, Position, ID}.
Actors will periodically send heartbeat messages to their neighbors to ensure that they are functional and also report changes to the one-hop neighbors.

1.2) Detecting Actor Failure and Initiating the Recovery Process:

Missing heartbeat messages can be used to detect the failure of actors. Depending on the actor's position in the network topology, major or no recovery may be needed. We focus on restoration of inter actor connectivity when a cut-vertex node fails. Neighbors of failed actors trigger the execution of DARA 1-C in localized manner. The failed actor is referred to thereafter as $A_f$.

1.3) Best Candidate Selection:

DARA-1C restores the connectivity of a partitioned network by substituting $A_f$ with one of its one-hop neighbors. The obvious question is which neighbor should be picked. DARA-1C strives to identify the BC for replacing $A_f$ using the following criteria in order:

1. Least node degree. Moving a node with larger number neighbors has greater impact on network. DARA-1C favors replacing the failed actor with the neighbor that has the least node degree.
2. Closest proximity to failed actor. In order to minimize the movement overhead, the nearest actor to $A_f$ will be favored.
3. Highest actor ID. It is possible that among the neighbors of Af , two or more actors have identical node degrees and are equidistant to it. The actor with the greatest ID will be picked to break the tie.

These three criteria guarantee that the same BC is identified at all neighbors of Af.

1.4) Cascaded Node Relocation:

The BC actor will prepare itself to move to the location of $A_f$ and calculate the expected time it will take to reach the new location. In addition, before moving to the new location, the BC will inform all its neighbors about its movement and the time it will take to reach to the new location by sending a "MOVING" message. The BC will then broadcast a "RECOVERED" message upon arriving at the destination.
The dependent neighbors (children) of the BC keep waiting until they receive the "RECOVERED" message indicating the restoration process has been completed and that they are still connected. Such scenario happens when the relocated actor stays in the radio range of these dependent children. If some of the dependents do not hear the "RECOVERED" message, they will assume that they got disconnected and apply DARA-1C again as if their parent has stopped functioning. In other words, the recovery process will be applied recursively to trigger the cascaded relocation of affected actors. Thus, these detached dependents identify a BC at the children level to relocate to the position of their parent. Please note that the child BC will do exactly what its parent has done, i.e., broadcast "RECOVERED" message to its neighbors when it ceases motion. This process continues until every dependent child is connected.

2) *DARA 2-C*

This section focuses on restoring 2-connectivity after the failure of an actor. When a critical actor fails, some of the nodes may be temporarily isolated until the network connectivity is restored, and thus, the network operation may be disrupted during the recovery. A 2-connected network maintains two independent paths among each pair of nodes. By This way we would ensure continual inter actor coordination even if an actor fails. Such robust operation is necessary since we made real time decisions distain WSANs.
DARA-2C is a completely distributed and localized restoration mechanism that can work in real time. The main is to recover from the failure of a boundary node. When a actor failed, each of the neighbours run DARA 2-C, since each node maintained at least two neighbours. The following are detailed steps of DARA 2-C.

2.1) Detecting Failure Nodes:

Similar to DARA-1C, actors will periodically send heartbeat messages to their neighbours to ensure that they are functional. Missing heartbeat messages can be used to detect the failure of actors.

2.2) Selecting node for reposition:

DARA-2C restore network connectivity with the neighbours of A that are boundary nodes to each other. DARA-2C sets selection criteria in order best candidate to move. The main criterion for picking the BC is the following:

1. Lowest node degree: The node that has the least number of neighbours will result few cascaded relocations.
In case multiple candidates have the least node degree, the following criterion is employed to qualify the best choice.
2. Least distance. In order to minimize the travel overhead, the closest candidate to A f, among those having the least node degree, If there is still tie the following criteria will be applied
3. Highest actor ID. The node that has greatest ID will be selected to break the tie.

2.3) Node Relocation

The BC actor, $A_{BC}$ will notify all its neighbours that it is moving and tell them where it intends to reposition. Similar to DARA-1C, when $A_{BC}$ moved, it broadcasts a ''RECOVERED'' message indicating the completion of the restoration process. The neighbours of $A_{BC}$ will keep waiting for the "RECOVERED" message; if they receive it, they conclude that they are still connected.

3) *Least Disruptive Topology Repair.*

LeDiR [5] is localised distributed algorithm which restores network connectivity with minimum number of node movements and ensures that no path length extended between any pair of nodes prior to failure. When a node fails, its neighbours will individually consult their possibly incomplete routing table to take appropriate course of actions and define their role in the recovery if any. If the failed node is critical node whose failure causes the network to partition into disjoint blocks, the neighbour that belongs to the smallest block reacts.
The main idea for LeDiR is to pursue block movement instead of individual nodes in cascade. To limit the travelled distance, LeDiR identifies the smallest among the disjoint blocks.
The following are the major steps:

3.1) Failure detection:

Actors will periodically send heartbeat messages to their neighbours to ensure that they are functional, and also report changes to the one-hop neighbours. Missing heartbeat messages can be used to detect the failure of actors. Once a failure is detected in the neighbourhood, the one-hop neighbours of the failed actor would determine the impact that is , whether the failed node is critical to network connectivity. This can be done using the shortest routing path table (SRT )by executing the well-known depth-first search algorithm.

3.2) Smallest block identification:
LeDiR limits the relocation to nodes in the smallest disjoint block to reduce the recovery overhead. The smallest block is the one with the least number of nodes and would be identified by finding the reachable set of nodes for every direct neighbour of the failed node and then picking the set with the fewest nodes. Since a critical node will be on the shortest path of two nodes in separate blocks, the set of reachable nodes can be identified through the use of the SRT after excluding the failed node. In other words, two nodes will be connected only if they are in the same block.

3.3) Replacing faulty node:

If node J is the neighbour of the failed node that belongs to the smallest block, J is considered the BC (Best candidate) to replace the faulty node. Since node J is considered the gateway node of the block to the failed critical node. The reason for selecting J to replace the faulty node is that the smallest block has the fewest nodes in case all nodes in the block have to move during the recovery.

3.4) Children movement:

When node J moves to replace the faulty node, possibly some of its children will lose direct links to it. In general, we do not want this to happen since some data paths may be extended. LeDiR opts to avoid that by sustaining the existing links. Thus, if a child receives a message that the parent P is moving, the child then notifies its neighbors (grandchildren of node P) and travels directly toward the new location of P until it reconnects with its parent again. If a child receives notifications from multiple parents, it would find a location from where it can maintain connectivity to all its parent nodes by applying the procedure used in RIM [4]. Briefly, suppose a child C has two parents A and B that move toward the previous location of node J. As previously mentioned, node J already moved to replace the faulty node F, and as a result, nodes A and B get disconnected from node J. Now, nodes A and B would move toward the previous location of J until they are r/2 units away. Before moving, these parents inform the child C about their new locations. Node C uses the new locations of A and B to determine the slot to which it should relocate. Basically, node C will move to the closest point that lies within the communication ranges of A and B, which is the closest intersection point of the two circles of radius r and cantered at A and B, respectively.

IV. COMPARISONS

Fig. 1 illustrates the difference between three algorithms. LediR restores network connectivity after failed actor $A_{14}$ by moving $A_{17}$ from the smallest block. RIM would stretch the links from both disjoint blocks and $A_{11}$ and $A_{17}$ toward $A_{14}$ to reconnect the network. The cascaded relocation for either LeDiR or RIM would not increase any shortest path. In case of actor $A_{11}$ will replace $A_{14}$ and actors $A_{12}$, $A_2$, $A_{13}$ would move during cascade relocation. As a result path length gets increased (e.g path length from $A_3$ to $A_{17}$ ). In DARA the length of the shortest path may grow as the parent moves to replace faulty node child also follow in the same path. In both RIM and DARA path length extended when network grows. The increase in path length will cause packet loss and data delivery delay. LeDiR strives to avoid that problem.
Table I provides a comparison of the analytical performance. As indicated in the table, LeDiR outperforms both baseline approaches when considering the recovery overhead at the network level in terms of the number of nodes participating in the recovery and the distance that these nodes collectively travel. RIM matches with LeDir in terms travel overhead. LeDiR guarantees minimum path length when compared to RIM and DARA.

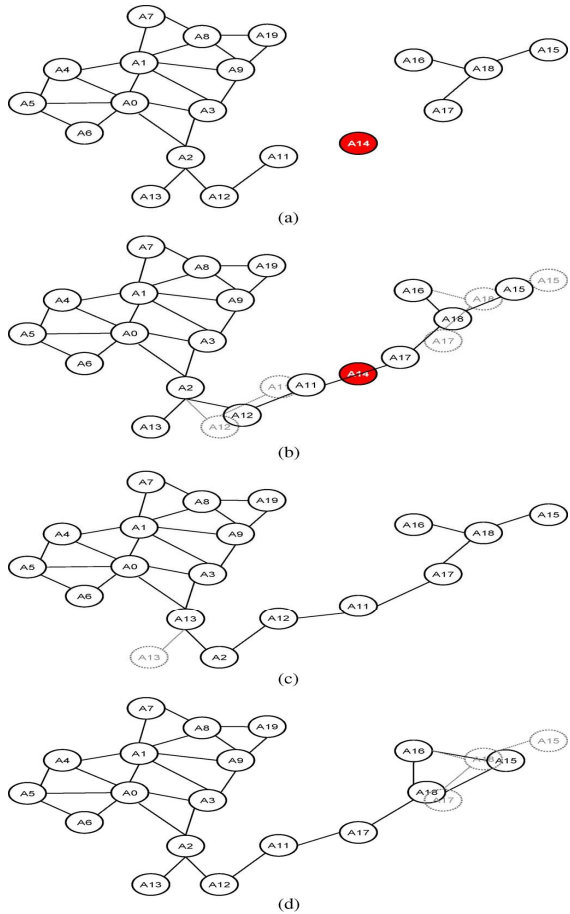

Fig.1. a) WSAN 1-connected topology with faulty node b)Topology recovered using RIM c) Topology recovered using DARA d) Topology recovered using LeDiR.

The following metrics are used to measure the performance in terms of recovery overhead.
1)Total travelled distance: reports the distance that the involved nodes collectively travel during the recovery. This can be envisioned as a network-wide assessment of the efficiency of the applied recovery scheme.
2) Number of relocated nodes: reports the number of nodes that moved during the recovery. This metric assesses the scope of the connectivity restoration within the network.
3) Number of exchanged messages: tracks the total number of messages that have been exchanged among nodes. This metric captures the communication overhead.
Furthermore, the following metrics are used to validate the path length performance :

1) Number of extended shortest paths: reports the total number of shortest paths between pairs of nodes (*i, j*) that get extended as a result of the movement-assisted network recovery..
2) Shortest paths not extended: reports average number of shortest paths that are not extended per topology:

| Property | LeDir | RIM | DARA |
|---|---|---|---|
| Maximum number of nodes to be involved | $\lfloor 1/2(N-1) \rfloor$ | N-1 | N-3 |
| Maximum messages to be sent | $\lfloor 3/2(N-1) \rfloor$ | 2N-1 | 5N-3 |
| Maximum distance travelled by a node | r | r/2 | r |
| Maximum Distance travelled by all engaged nodes | $\lfloor r/2(N-1) \rfloor$ ≈1/2 r N | $\lfloor r/2(N-1) \rfloor$ ≈1/2 r N | r(N-3)≈rN |

Table 1. Analytical performance of LeDiR, RIM, and DARA. Where N represents number of deployed nodes and r represents communication range

| Performance metrics | Dense Topologies | | |
|---|---|---|---|
| | LeDiR | RIM | DARA |
| Network overhead | Outperforms over RIM | well | ---- |
| Path length validation | Excellent | Poor(cont Tolerated) | Poor(cont Tolerated |

Table 2. Performances analysis in Dense Topologies

| Performance metrics | Sparse Topologies | | |
|---|---|---|---|
| | LeDiR | RIM | DARA |
| Network overhead | Equal to RIM | well | ---- |
| Path length validation | Excellent | Poor(con be tolerated) | Poor(can be Tolerated) |

Table 3. Performances analysis in Dense Topologies

Overhead related metrics: as we see in Table 2 LediR outperforms over RIM in network overhead related metrics in terms number of nodes moved, distance travelled and number of exchanged messages. Where in sparse topologies given in Table 3 both LeDiR and RIM have equal performance
Path length validation metrics: LeDiR does not extend any shortest paths unlike RIM and DARA. This cannot be tolerable in dense topologies as we see in table 2. Where as in sparse topologies given in table 3 path length extensions is somehow tolerable.

## V. CONCLUSION

In recent years, wireless sensor and actor (actuator) networks (WSANs) have been deployed in harsh environments where human intervention is almost difficult to take place. Failure of actors in such areas partition the network into disjoint sub sets. This paper dealt with review of three failure node recovery algorithms in order to re-establish network connectivity after node failure. We have been analyzed performance of these three algorithms with respect to network overhead and path length validation metrics.